\documentclass[longbibliography,amsmath,amssymb,prl,twocolumn,showpacs,final]{revtex4-2}
\usepackage{graphicx} 

\usepackage{xcolor}
\usepackage{oplotsymbl}
\definecolor{color-1}{rgb}{0,0,1}
\definecolor{color-2}{rgb}{0.15,0.15,0.15}

\begin{document}

\definecolor{c1}{HTML}{440154}
\definecolor{c2}{HTML}{33638D}
\definecolor{c3}{HTML}{FDE725}

\title{Self-Diffusion in Inhomogeneous Granular Shearing Flows}

\author{Riccardo Artoni}
\email{riccardo.artoni@univ-eiffel.fr}
\affiliation{MAST-GPEM, Université Gustave Eiffel, IFSTTAR, F-44344 Bouguenais, France}%

\author{Michele Larcher}
\affiliation{Free University of Bozen-Bolzano, I-39100 Bozen-Bolzano, Italy}%

\author{James T. Jenkins}
\affiliation{Cornell University, Ithaca, NY 14053, USA}%
\author{Patrick Richard}
\affiliation{MAST-GPEM, Univ Gustave Eiffel, IFSTTAR, F-44344 Bouguenais, France}%
\date{\today}

\begin{abstract}
In this letter, we discuss how flow inhomogeneity  affects the self-diffusion  behavior in granular flows. Whereas self-diffusion scalings have been well characterized in the past for homogeneous shearing, the effect of shear localization and nonlocality of the flow has not been studied. We therefore present measurements of self-diffusion coefficients in discrete numerical simulations of steady, inhomogeneous, and collisional shearing flows of nearly identical, frictional, and  inelastic spheres. We focus on a  wide range of  dense solid volume fractions,  that correspond to geophysical and industrial shearing flows that are dominated by collisional interactions.  We compare the measured values first, with a scaling based on shear rate and, then, on a scaling based on the granular temperature. We find that the latter does much better than the former in collapsing the data. The results lay the foundations of diffusion models for inhomogeneous shearing flows, which should be useful in treating problems of mixing and segregation.
\end{abstract}
                              

\maketitle

\paragraph{Introduction.} Diffusion is a key mechanism for mass transfer in granular shearing flows. In such flows, collisions between spheres induce velocity fluctuations of the grains that drive the motion of particles in a way that is analogous to the thermal diffusion in  dense gases, or the dispersion induced by eddies in turbulent flows. 
Such shear-induced diffusion is important for applications in industrial and natural granular flows involving mixing and segregation. The problems encountered share many features with  other research involved with transport phenomena and motion of discrete bodies, and therefore advances in the comprehension of diffusion are profitable to several soft matter fields (granular materials, suspensions, traffic and crowd dynamics, active matter, ...). 
Experimental and numerical studies have been devoted to characterize diffusion in various flow geometries, often in connection with segregation: granular shear cells or simple shear flows \cite{scott76,buggisch89, savage93}; vertical channels \cite{hunt92,natarajan95};  inclined chutes \cite{zik91}; rotating drums \cite{khan05, taberlet06}.
Phenomenological theories of segregation during flow  (e.g. Refs. \cite{gray11,fan11}), that produce plausible predictions of species’ concentrations and mixture velocity for appropriate choices of parameters,  generally involve simple models for  the mechanism of diffusion.

Despite the amount of studies, few have addressed in detail the anisotropic nature of self-diffusion and its scaling with the relevant kinematic parameters. 
For dilute systems (solid fraction lower than 0.5), Campbell \cite{campbell97} measured the self-diffusivity tensor by means of numerical simulations of a homogeneously sheared system of frictional, inelastic spheres. 
Utter and Behringer \cite{utter04} performed experiments  on slow, rate-independent, and inhomogeneous shearing of a dense aggregate of disks in a Couette cell, pointing to the existence of anisotropy and showing that in the dense limit, self-diffusion coefficients scale approximately as $D=kd^2\dot\gamma$, where $\dot\gamma$ is the shear rate, and $d$ is the particle diameter. It was later shown that in three dimensions, $k$ is a constant  on the order of 0.05~\cite{artoni21,fry19, cai19}. 
More recently, we~\cite{artoni21} extended Campbell's work by numerically studying dense systems (solid volume fraction in the range of 0.49 to 0.587) under homogeneous shear, characterizing the full self-diffusion tensor and discussing different scalings for the transverse diffusivity. For the same simplified flow configuration, Macaulay and Rognon \cite{macaulay19} investigated the effect of inter-granular cohesive forces on diffusion. 

Nearly all these studies dealt with homogeneous shearing. Inhomogeneous shearing flows are obviously more complex. In regions of shear localization and creep, and those influenced by boundaries~\cite{artoni18}, the rheology is known to become nonlocal, which means that the local effective friction is not simply determined by a balance between shear and confinement. In analogy with molecular theories, the introduction of velocity  fluctuations as an additional variable seems a promising path. 
The strength of velocity fluctuations is a classical ingredient of diffusion theories, and dense kinetic theories for the segregation of binary mixtures of inelastic spheres \cite{jenkins89,arnarson98,arnarson04,larcher13,larcher15} predict diffusion coefficients that exhibit explicit dependence on the volume fraction, the strength of the particle velocity fluctuations and the particles'{} size, mass, and collision properties. 
The important question we  address here in detail is how the inhomogeneity of the flow affects self-diffusion, and, in particular, whether the scaling of diffusivity on shear rate has to be replaced by one based on granular temperature, inspired by kinetic theory \cite{artoni21}. 

Here, we focus on flows in the range of volume fractions between 0.35 and 0.6. This comprises the range of volume fraction important in geophysical flows on Earth. Above a volume fraction of 0.49, long range order can appear in an equilibrated system of identical elastic spheres  \cite{alder57}, and at a volume fraction of about $\phi_c=0.587$, collisional flow becomes impossible, when the coefficient of sliding friction is 0.5  \cite{berzi15}. 
In order to address the question raised above, we measure  the components of self-diffusion parallel and  perpendicular to the flow  in discrete numerical simulations of inelastic spheres in dense, steady, and inhomogeneous shearing flows. We test the scaling based on the shear rate with that based on the granular temperature, and determine that the kinetic theory scaling does far better than the shear rate scaling in collapsing the data for inhomogeneous flows over a range of very dense volume fractions and strength of inhomogeneity.

\paragraph{Flow configurations.}

Discrete element method simulations were performed by means of the open source LAMMPS platform~\cite{lammps}. The numerical samples were slightly polydisperse, characterized by uniform (number-based) particle size distribution  in the range of $0.9d-1.1d$, where $d$ is the characteristic particle diameter. Masses were nondimensionalized with respect to the mass $m$ of a particle with size $d$ (in practice this corresponds to take a normalized density of $6/\pi$). Simulations were performed in a cuboidal, fixed volume cell ($L=20d$). A classical linear spring-dashpot model, with tangential elasticity and friction, was chosen as a contact model. The viscous  damping coefficient was adjusted to yield a normal coefficient of restitution $e_n=0.7$, and the interparticle friction coefficient was taken equal to $\mu=0.5$. Tangential stiffness was taken proportional to the normal one $k_t/k_n=2/7$, a classical assumption which ensures that the period of tangential oscillations is the same as the period of normal oscillations after a contact.

\begin{figure}[t]
\includegraphics[height=0.35\columnwidth]{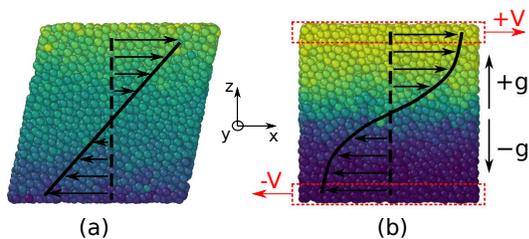} 
\caption{ The two flow configurations studied in this Letter: (a) homogeneous shear flow and (b) concave flow.  Cell height is $20d$. The lighter yellow corresponds to the largest velocity to the right, and the darker blue corresponds to the largest velocity to the left.}\label{fig:1}
\end{figure}

Results from two types of flows (displayed in Fig. \ref{fig:1}) are presented in the following. The flows are steady, fully-developed and quasi-bidimensional. For both flow configuration, we identify $x$ as the flow, or streamwise, direction, $z$ as the direction in which we impose a velocity gradient, and $y$ the transverse direction. In terms of the mean velocity field, $\vec v=(v_x(z), 0,0) $.
The first configuration is homogeneous shearing without gravity, which was the subject of a previous work~\cite{artoni21}, and which is taken here as the reference case. Homogeneous shear was obtained by means of fully periodic boundary conditions coupled with continuous domain deformation, as in LAMMPS's `fix deform' scheme. The solid fraction was varied in the range of $0.49-0.587$ in this configuration. Given that the volume was fixed, this was simply obtained by tuning the number of particles $N$.
The second configuration may be referred to -  after the shape of the shearing profile - as a concave flow \cite{kim20}. The shearing profile results from a nontrivial pattern of external forces given by: $F_z =  m g \frac{z - z_0}{|z-z_0|}$ where $z_0$ is the midpoint of the system. The shearing takes place at constant volume, by imposing the velocity ($\pm V$) of two layers at the extremes of the cell in the $z$-direction, each layer  being $2d$ thick (the effective flow height being therefore $H=16d$). Periodic boundary conditions hold in the other directions. In practice, the global shear rate (i.e. $2 V /H$) was kept constant and the gravitational acceleration $g$ was varied in order to vary the $z$-heterogeneity of the system. The initial state of the system was extracted from homogeneous shear tests at three different global solid fractions ($\phi\approx 0.53, 0.55, 0.57$).

The systems studied possess  three time scales:  one related to contact stiffness $\tau_s = \sqrt{m/k_n}$, one related to shearing $\tau_\gamma = 1/\dot\gamma$ and  one related to pressure $\tau_p = d /\sqrt{p/\rho}$. 
The latter, $\tau_p$ and $\tau_\gamma$ have to be much larger than the former, $\tau_s$, to keep the particle contacts sufficiently rigid. This guided the choice of the shear rate, which was chosen in order to ensure  $\tau_\gamma \sim 10^3 \tau_s $ locally for the homogeneous shear, and globally for the concave flow simulation (where it was verified that locally, even for the larger shear rates, $\tau_\gamma > 2\times 10^2 \tau_s $).  The condition $\tau_p>> \tau_s $ was also verified.
As is well known, the time scales relative to shearing and pressure can be combined into a dimensionless number, called the inertial number, which is sometimes used to characterize the flow regimes : $I= \frac{\tau_p}{\tau_\gamma}=\frac{\dot\gamma d}{\sqrt{p/\rho}} $. In the concave flow configuration, a global estimate considering all the acceleration values used ($g\sim 10^{-7}-10^{-5}\frac{dk_n}{m}$) yields $I_{glob} = \frac{Vd}{H\sqrt{g H}} =0.14-1.4$ for the range of parameters used in this Letter. Considering the local shear rate (which is in principle different from $2 V/H$ due to the heterogeneity of the flow), yields a  range of local inertial numbers $I_{loc}=4\times 10^{-2} -5$, which means that our data are relevant for dense-to-rapid granular flows. Please note that during preparation, the material was subject to a global cumulative deformation of about $\dot\gamma\Delta t  \sim 10^4$ (locally $> 10^3$). We indeed checked that this is enough to obtain a system at stationary state. 

\begin{figure}[t]
\includegraphics[width=0.99\columnwidth]{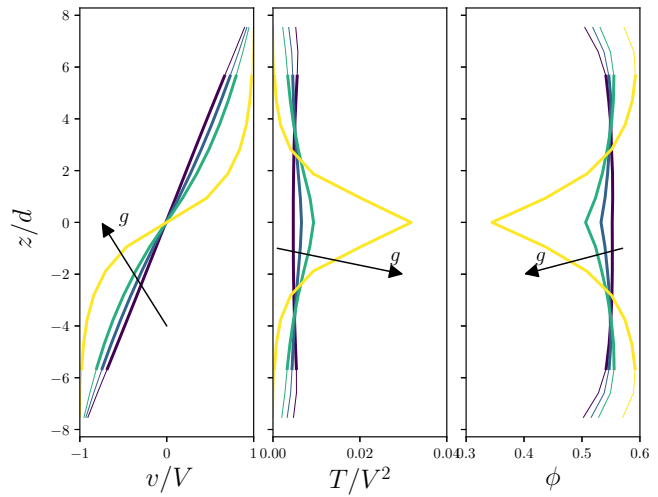}
\caption{ Average flow profiles in the concave flow, for a global value of solid fraction of $\phi\approx0.55$ and different values of acceleration $g$ (respectively $ g= 10^{-7}, 10^{-6}, 2\times 10^{-6},3\times10^{-6}\frac{dk_n}{m} $ when going from dark blue to yellow ): (a) streamwise velocity, (b) granular temperature, (c) solid fraction. The part of the profiles in bold corresponds to data points which were used for the self-diffusion analysis.}\label{fig:2}
\end{figure}

\paragraph{Inhomogeneous flow profiles.}
In order to highlight the heterogeneity of the concave flow configuration, we display in Fig.~\ref{fig:2} the profiles of average streamwise velocity, granular temperature and solid fraction. Averages were performed on layers of thickness equal to $1d$, and velocity fluctuations were computed with a correction based on a gradient expansion as in ref. \cite{artoni15}. The profiles in the figure correspond to a global solid fraction of $\phi\approx0.55$ and different values of the acceleration $g$.
It is clear that the velocity profile becomes S-shaped with increasing $g$. A zone of higher shear appears, therefore, in the middle of the sample, characterized also by higher granular temperature and lower solid fraction. The central zone becomes, therefore, increasingly collisional and dilute as $g$ increases, whereas, on the other hand, the peripheral zones get denser. Note that above a certain value of $g$ (which depends on the average solid fraction, or $N$), the peripheral zones become so dense that the central zone is empty of particles. The system is then composed by two separate solid blocks (and is of no more of interest). This sets a limit to the operational range of $g$.
We can conclude that the introduction of a complex force pattern introduces a strong heterogeneity of the kinematic profiles and, particularly, yields gradients of granular temperature. These gradients are strong (i.e. $d\nabla T / T\sim 1$)  in the most heterogeneous cases.
Granular temperature heterogeneity is often associated with the need of using nonlocal rheologies~\cite{zhang17}, and  the concave flow configuration has been introduced to study such nonlocal effects~\cite{kim20}. The concave flow configuration is also an interesting case for the understanding of how self-diffusivity scalings may be affected by such nonlocal effects.

\paragraph{Self-diffusivity computation.}
As in previous works, we characterise the self-diffusivity from the trajectories of the particles via the mean squared displacements. For concave flows, given that the flows are inhomogeneous, the analysis is performed on layers one-diameter in thickness, normal to $z$. For each layer, the part of particle  trajectories belonging to the layer are considered and averaged to characterize the time scaling of the mean squared displacements via the formula:
\begin{equation}
\left\langle \Delta  x_i \Delta x_j\right\rangle \propto \Delta t^{\alpha },
\end{equation}
where an exponent $\alpha$ close to unity is related to a diffusive behavior. As it was previously shown, for short times $\alpha >1$, because particle velocity correlations yield a ``super-diffusive'' behavior. In order to characterize the long-time behavior, we analyze displacements for time lags larger than the correlation time. In particular, we use an approximation for the correlation time based on granular temperature, and therefore  we analyze displacements for rescaled time lags $\Delta t \frac{\sqrt T}{d} > 1$.
On the other hand, the length of the trajectories is not constant (and bounded) because the particles can exit the layers; therefore, for long times there is not enough statistics. The local granular temperature which measures the intensity of fluctuations,  is relevant in giving a timescale for this problem. We have seen that, in our numerical setup, enough statistics is available for rescaled time lags $\Delta t \frac{\sqrt T}{d} <15$. Within the  identified range of time lags, a power law fit permits the verification of the diffusive behavior and the computation of the diffusivity.

Note, however, that the subdivision of the system in thin layers limits the maximum observable displacement in the $z$ direction, which means that the diffusivity components related to that direction (i.e.  (zz), (xz) and (yz)) cannot be computed by the method.
Details on the time scaling of the mean squared displacements with time is given in the Supplementary Material \ref{supp}. Here we limit ourselves to mentioning the main results. First of all, the transverse, (yy), component of the mean squared displacement is clearly diffusive ($\alpha \approx 1$), whereas the mixed (xy) appears to be close to zero. 
Regarding the streamwise (xx) component, in order to compute the displacements, the part of the displacement associated with the shear flow was subtracted from the total displacement in order to avoid considering Taylor dispersion effects.

As detailed in the SM, two types of averaging can be adopted to determine the average shear flow: (i) instantaneous spatial and (ii) time spatial averaging. We find that the type of averaging influences the observed scaling: when subtracting the average velocity profile determined by global time-space averaging, the mean square displacements may display super-diffusive behavior. The observed super-diffusivity is stronger  where temperature gradients are larger ($d\nabla T/ T \rightarrow 1$).  On the other hand, using an instantaneous average velocity profile to correct the displacement yields a diffusive behavior. 

Given that super-diffusion nearly disappears when changing the definition of fluctuations, it seems reasonable to postulate that its origin has to be found in  multi-particle, ``eddy-like'' time correlations which are removed  by instantaneous averages. This can be put in relation to the debated concept of ``granulence'' and the findings of super-diffusion in quasistatic flows~\cite{radjai02}, and to the concept of granular fluidity, which has been related to  velocity fluctuations defined with respect to instantaneous averages~\cite{zhang17}.
Our results support the idea that it may be  important to analyze separately the fluctuations at the scale of a particle from those at the scale of several particles. 

\paragraph{Self-diffusion scaling.}

\begin{figure}[h]
\includegraphics[width=0.99\columnwidth]{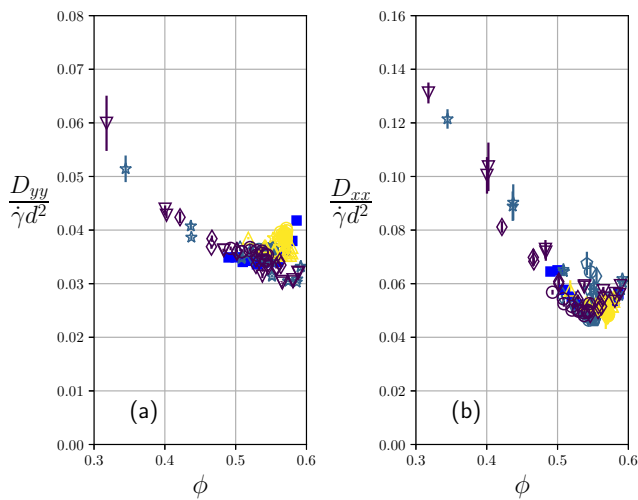}
\caption{(a) Transverse ($D_{yy}$) and (b) streamwise  ($D_{xx}$) self-diffusivities, as determined from numerical simulations, scaled by shear rate, vs solid volume fraction. The blue filled  squares (\textcolor{blue}{$\squadfill$}) correspond to homogeneous shearing, and the empty symbols correspond  to concave flow. Colors correspond to different average solid fraction values: \textcolor{c1}{0.53}, \textcolor{c2}{0.55}, \textcolor{c3}{0.57},  and symbols to different values of $\frac {g m } {d k_n}   =  10^{-7} (\pentago),  10^{-6}(\circlet),  2\times 10^{-6} (\rhombus),3\times 10^{-6} (\trianglepb),5\times 10^{-6}(\starlet),$ and $ 10^{-5}(\trianglepa)$ . Error bars correspond to the 95\% confidence interval of the diffusivities determined from the linear fit of the respective averaged mean squared displacement .}\label{fig:6}%
\end{figure}

Based on the above discussion, it is evident that in concave flow, it is possible to evaluate a self-diffusivity coefficient  (yy)  for the transverse direction, and  (xx) for the streamwise direction only when the particle displacements are corrected with the instantaneous spatial velocity gradients.  These diffusivities are computed by fitting the mean squared displacement data with the formula:
\begin{equation}
\left\langle \Delta x_i^2\right\rangle  = 2D_{ii}\Delta t.\label{Dyy}
\end{equation}    

We  study scalings of $D_{xx}$ and $D_{yy}$ as a function of system parameters and in comparison to previous homogeneous shear results \cite{artoni21}.
Following a scheme discussed in our previous work, we check the relevance of the simplest empirical scaling, based on shear rate; in other words, we study the functions $\frac{D_{ii}}{\dot\gamma d^2}$ vs $\phi$. These scalings are displayed in Fig.~\ref{fig:6}(a) and \ref{fig:6}(b), where homogeneous shear flow results for $D_{xx}$ and $D_{yy}$ are also reported for comparison.
A first general comment is that, in inhomogeneous shear flows, the hierarchy of self-diffusivities is the same as in homogeneous shear ($D_{xx}>D_{yy}$) and, as is well known~\cite{campbell97}, the self-diffusivities increase (and the difference between $D_{xx}$ and $D_{yy}$ increases as well) when $\phi$ decreases  below $0.5$.
There is also a global qualitative agreement between homogeneous shear flow and the concave flow results; however, important deviations  are evident, particularly for the more heterogeneous flows (e.g. simulations with large $g$, denoted by symbols $\starlet$ and $\trianglepb$ in Fig.~\ref{fig:6}). The empirical scaling based on shear rate, which can be thought as a ``local'' scaling, while providing a correct order of magnitude, appears to fail in predicting the effect of temperature gradients and its validity seems, therefore, limited to nearly homogeneous flows ($d\nabla T/ T << 1$).  In heterogeneous flows, a more refined analysis may be needed. Moreover, for the (xx) component, the emergence of superdiffusivity related to eddy-like correlations has to be characterized in detail, given that it also determines the macroscopically observable diffusion. 

\begin{figure}[h]
\includegraphics[width=0.99\columnwidth]{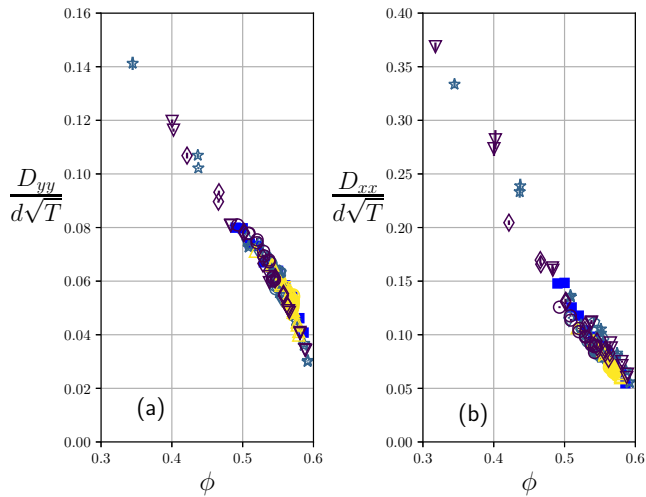}\caption{(a) Transverse ($D_{yy}$) and (b) streamwise  ($D_{xx}$) self-diffusivities, as determined from numerical simulations, scaled by the square root of granular temperature, versus solid volume fraction. The symbols have the same significance as in Fig. \ref{fig:6}.}\label{fig:7}
\end{figure}

Here we limit ourselves to test whether a scaling based on the granular temperature (often associated with ``nonlocal'' models of granular flows~\cite{berzi2015,berzi2020}) allows better representation of the self-diffusivity results, encompassing also the most heterogeneous simulation data. 
The  scalings  $\frac{D_{ii}}{d\sqrt{T}}$ vs $\phi$ are shown in Fig.~\ref{fig:7}(a) and \ref{fig:7}(b). It is evident that the concave flow data superpose (better here than in Fig.~\ref{fig:6}) to those of homogeneous shearing. 
As we showed in our previous work dealing with homogeneous shear flow~\cite{artoni21}, even in this case the kinetic theory of granular gases~\cite{brilliantovbook} underestimates  the transverse self-diffusivity by a factor increasing with $\phi$ and reaching a value of about 3 for denser systems. 
Nevertheless, the physically based self-diffusivity scaling  on granular temperature behaves remarkably better than the empirical scaling on the shear rate, for both the transverse and the streamwise components, and this is so particularly when temperature gradients are important.   Our data may, therefore, be used to extend theoretical predictions for the self-diffusivity tensor, which were developed for dilute flows~\cite{garzo02,abbas09}, to dense systems.
 
\paragraph{Conclusions.}
We have measured self-diffusion coefficients in discrete numerical simulations of both homogeneous and inhomogeneous shearing flows of nearly identical frictional and inelastic spheres. The studied systems were relatively dense, {i.e.}, solid fractions from approximately $0.35$ to approximately $0.59$. In homogenous flows the diffusivity can be estimated by using the grain size $d$ and the inverse of the shear rate $1/\dot\gamma$ as, respectively, the natural length and time scales, {i.e.,} $\dot\gamma$: $D\propto d^2 \dot\gamma$. Yet, this purely local scaling, does not hold for inhomogeneous flows for which non-local effects are important. Here we measured  $D_{xx}$ and $D_{yy}$, i.e.  the streamwise and the transverse diffusivities, highlighted their anisotropy ($D_{xx}>D_{yy}$), and showed that the scaling $D_{ii}/d\sqrt{T}$ with $T$ as the granular temperature leads to a better description of the evolution of the transverse diffusivity with the solid fraction than $D_{ii}/\dot\gamma d^2$.  Besides explaining diffusive properties in inhomogeneous granular shearing flows, this feature could help understanding the non-local behavior of granular flows. Our results are, therefore,  of immediate interest for researchers working on transport phenomena in dense shearing flows of discrete entiies, such as suspensions, traffic flow, crowd dynamics, and active matter.

\begin{acknowledgments}
\paragraph{Acknowledgments.}  Authors  RA and PR acknowledge financial support from ANR (Grant No. ANR-20-CE08-0028 MoNoCoCo). The numerical simulations were carried out at the CCIPL (Centre de Calcul Intensif des Pays de la Loire) under the project ``Simulation numérique discrète de la fracture des matériaux granulaires''.
\end{acknowledgments}


\end{document}